\documentclass{aadebug}
\corr{0309027}{71}

\begin{document}

\runningheads{Bryan Cantrill}{Postmortem Object Type Identification}

\title{Postmortem Object Type Identification}

\author{
Bryan~M.~Cantrill\addressnum{1}
}

\address{1}{
Sun Microsystems,
17 Network Circle,
Menlo Park, California
}

\pdfinfo{
/Title (Postmortem Object Type Identification)
/Author (Bryan Cantrill)
}

\begin{abstract}
This paper presents a novel technique for the automatic type identification of
arbitrary memory objects from a memory dump.  
Our motivating application is debugging memory
corruption problems in optimized, production systems --- a problem domain
largely unserved by extant methodologies.  We describe our algorithm as
applicable to any typed language, and we discuss it with respect to the
formidable obstacles posed by C.  We describe the heuristics that
we have developed to overcome these difficulties and achieve effective type
identification on C-based systems.  We further describe the implementation of
our heuristics on one C-based system --- the Solaris operating system
kernel --- and describe
the extensions that we have added to the Solaris postmortem debugger to allow
for postmortem type identification.  We show that our implementation yields a
sufficiently high rate of type identification to be useful for debugging
memory corruption problems.  Finally, we discuss some of the novel automated
debugging mechanisms that can be layered upon 
postmortem type identification.
\end{abstract}


\section{Introduction}

While there are a myriad of different techniques for automatically
debugging memory
corruption problems, they share one conspicuous trait:  each induces a
negative effect on run-time performance.  
In the least invasive techniques the effect is merely moderate,
but in many it is substantial --- and in none is the performance effect
so slight as to allow the technique to be enabled at all times in
production code.
As such, the applicability of these techniques is limited to 
{\em reproducible} memory corruption problems in {\em development}
environments.
These techniques are ineffective for debugging the most
virulent memory
corruption problems: {\em non-reproducible} problems in
{\em production} environments.
The only data from such problems is the state of the system
itself: when memory corruption induces a fatal failure in the system, a
snapshot of state is typically taken and copied to stable storage.
To be applicable to these memory corruption 
problems, automatic debugging techniques must assume an optimized
system, and restrict themselves to making use only of {\em postmortem} state.  

\begin{figure}[t!]
\begin{center}
\begin{picture}(200,200)
\put(0, 180){
\put(0, 0){\makebox(70,20){{\tt ...}}}
\put(70, 0){\makebox(70,20){{\tt ...}}}
}
\put(0, 160){
\put(0, 0){\makebox(70,20){{\tt 0xde4ecd10}}}
\put(70, 0){\makebox(70,20){{\tt 0x29010601}}}
}
\put(0, 140){
\put(0, 0){\makebox(70,20){{\tt 0xde4ecd14}}}
\put(70, 0){\makebox(70,20){{\tt 0x25000002}}}
}
\put(0, 120){
\put(0, 0){\makebox(70,20){{\tt 0xde4ecd18}}}
\put(70, 0){\makebox(70,20){{\tt 0x06010201}}}
}
\put(0, 100){
\put(0, 0){\makebox(70,20){{\tt 0xde4ecd1c}}}
\put(70, 0){\makebox(70,20){{\tt 0x08010201}}}
}
\put(0, 80){
\put(0, 0){\makebox(70,20){{\tt 0xde4ecd20}}}
\put(70, 0){\makebox(70,20){{\tt 0x23000001}}}
}
\put(0, 60){
\put(0, 0){\makebox(70,20){{\tt 0xde4ecd24}}}
\put(70, 0){\makebox(70,20){{\tt 0xde76b734}}}
}
\put(0, 40){
\put(0, 0){\makebox(70,20){{\tt 0xde4ecd28}}}
\put(70, 0){\makebox(70,20){{\tt 0xdedfbc68}}}
}
\put(0, 20){
\put(0, 0){\makebox(70,20){{\tt 0xde4ecd2c}}}
\put(70, 0){\makebox(70,20){{\tt 0xde769afc}}}
}
\put(0, 0){
\put(0, 0){\makebox(70,20){{\tt ...}}}
\put(70, 0){\makebox(70,20){{\tt ...}}}
}
\multiput(70, 20)(0, 20){9}{
\multiput(0, 0)(5, 0){14}{\line(1, 0){2}}}

\put(70, 0){\line(0, 1){200}}
\put(140, 0){\line(0, 1){200}}
\put(142, 25){\oval(10, 10)[br]}
\put(147, 25){\line(0, 1){30}}
\put(152, 55){\oval(10, 10)[tl]}
\put(142, 95){\oval(10, 10)[tr]}
\put(147, 95){\line(0, -1){30}}
\put(152, 65){\oval(10, 10)[bl]}

\put(0, 80){
\put(142, 25){\oval(10, 10)[br]}
\put(147, 25){\line(0, 1){30}}
\put(152, 55){\oval(10, 10)[tl]}
\put(142, 95){\oval(10, 10)[tr]}
\put(147, 95){\line(0, -1){30}}
\put(152, 65){\oval(10, 10)[bl]}
}

\put(173, 61){\makebox(0, 0)[b]{\footnotesize{{\em Corrupted}}}}
\put(173, 60){\makebox(0, 0)[t]{\footnotesize{{\em buffer}}}}

\put(173, 141){\makebox(0, 0)[b]{\footnotesize{{\em Unknown}}}}
\put(173, 140){\makebox(0, 0)[t]{\footnotesize{{\em buffer}}}}

\end{picture}
\caption{\small{Example of buffer overrun memory corruption.
The first word
of the 16-byte buffer at address 0xde4ecd20 has been corrupted,
inducing fatal error when the memory stored there (0x23000001)
was interpreted as a pointer.  The
similarity between the corruption value and the contents of the 
buffer at address 0xde4ecd10 indicates that the code manipulating
the unknown buffer is likely
responsible for the corruption.}}
\label{figbuf}
\end{center}
\end{figure}

Memory corruption problems have several variants, but 
a particularly common pathology is a {\em buffer
overrun}, in which a memory object is erroneously treated as 
a memory object of larger size\cite{sullivan91software}.  Because
the errant subsystem stores to
the memory beyond the bounds of the object, this pathology can
induce wildly varying manifestations:  system failure is typically
induced not by the corrupting subsystem,
but rather by a disjoint (and
otherwise correct) subsystem that happens to have a memory buffer
adjacent to that of the errant subsystem.  
As Figure \ref{figbuf} shows, when debugging these problems postmortem
it is often apparent based on buffer contents that one buffer has
overrun the other.  
The subsystem associated with the 
victimized buffer is known by virtue of that subsystem having induced
the fatal error; in order to make progress, the 
subsystem associated with the errant buffer must be determined.
In a system that makes widespread use of derived types, determining the 
{\em type} of a buffer can often implicitly identify the subsystem
 --- and at the very least, determining the type considerably focuses the
search for errant code.

We have developed an automatic technique for determining
the type of an arbitrary memory object from a memory dump of an optimized
system.
As many optimized systems are implemented in C, we have
developed some specific heuristics to make our technique effective
on C-based systems.  We have implemented these heuristics on Solaris
for use in debugging Solaris kernel crash dumps, and have in practice been
able to achieve very high rates of type identification --- typically better
than 80 percent of all memory objects, and often better than 95 percent.
This high rate of identification has allowed our technique to be used to
successfully debug otherwise undebuggable kernel memory corruption problems.

The remainder of this paper is structured as follows:
Section~\ref{sec-related} discusses related work;
Section~\ref{sec-algorithm} discusses our technique in general;
Section~\ref{sec-problems} discusses the problems that a C-based system
poses with our technique; Section~\ref{sec-heuristics} discusses the
heuristics that we have developed to overcome the obstacles outlined in
Section~\ref{sec-problems}; Section~\ref{sec-implementation} discusses
the details of our implementation; Section~\ref{sec-otherapps} describes
other applications of postmortem type identification;
Section~\ref{sec-future} outlines
areas for future work.

\section{Related work}
\label{sec-related}

\begin{sloppypar}
There has been a substantial body of work devoted to debugging 
memory corruption problems in development, including
Purify\cite{purify}, Sabre-C\cite{saber},
Kendall's {\tt bcc}\cite{bcc},
Steffen's {\tt rtcc}\cite{Steffen:1992:ART}, and Jones and Kelly's bounds
checking\cite{jones97backwardscompatible}.
Many of these
require recompilation, and all of them induce substantial performance
impact, varying from as little as 130\%\cite{austin94efficient} to as much as
20,000\%\cite{saber}.  None of
these target production code explicitly, and Jones and
Kelly even conclude that it is ``unlikely that we could ever achieve the
10-15\% performance loss that would be acceptable if programs are
to be distributed with bounds checks
compiled in.''\cite{jones97backwardscompatible}
\end{sloppypar}

The problem of debugging memory corruption problems in production
was explicitly identified by Patil and Fischer in \cite{patil95efficient},
in which they describe using
idle processors to absorb their technique's substantial performance impact.
Unfortunately, this is not practical in a general-purpose system: idle 
processors 
cannot be relied upon to be available for extraneous processing.  
Indeed, in performance critical systems {\em any} performance impact is often
unacceptable.  

Some memory allocators have addressed debugging problems in
production by allowing their behavior
to be dynamically changed to provide greater debugging
support\cite{bonwick94slab}.
This allows optimal allocators to be deployed
into production, while still allowing their debugging features to be
later enabled
should problems arise.  
A common way for these allocators to detect buffer overruns is to
optionally place {\em red zones} around allocated
memory.  However, this only provides for immediate identification of the
errant code if stores to the red zone induce a synchronous
fault.  Such faults are typically achieved by coopting the virtual
memory system
in some way --- either by surrounding a buffer with unmapped regions,
or by performing a check on each access.  The first has enormous cost
in terms of space, and the second in terms of time --- neither can
be acceptably enabled at all times.  Thus, these approaches are still
only useful for {\em reproducible} memory corruption problems.

If memory corruption cannot be acceptably prevented in production code,
then the focus must shift to debugging the corruption postmortem.
While the notion of postmortem debugging has existed since the
earliest dawn of debugging\cite{GillS1951a}, there seems to have been
very little work on postmortem debugging of memory corruption per se;
such as it is, work on postmortem debugging
has focused on race condition detection in parallel and distributed
programs.  The lack of work on postmortem debugging 
is surprising given its clear advantages for debugging production systems --- 
advantages that were clearly elucidated by McGregor and Malone
in \cite{McGregor:1980:SDI}:

\begin{quote}
A major advantage of this method of obtaining information about a
program's malfunction is that there is virtually no runtime overhead in 
either space or speed.  No extra trace routines are necessary and the
dump interpreting software is a separate system utility which is only
used when required.  This is a facility which remains effective when
a program has passed into production use and is very effective
in `nailing' those occasional bugs in a production environment.
\end{quote}

The only nod to postmortem debugging of memory corruption seems to
come from
memory allocators such as the slab allocator\cite{bonwick94slab} used
by the Solaris kernel.  This allocator can optionally 
log information with each allocation and deallocation; in the event of
failure, these logs can be used to determine the subsystem allocating 
the overrun buffer.  
While this mechanism has proved to be enormously useful in debugging
memory corruption problems in the Solaris kernel, it is still far
too space- and
time-intensive to be enabled at all times in
production environments.  

\begin{figure*}[t!]
\begin{center}
\begin{picture}(465,225)

\put(-19,0){
\put(0, 0){
\put(22, 170){
\put(0, 25){\makebox(86,20)[b]{static ``{\tt foo\_list}''}}
\put(0, 0){\framebox(86,20){{\tt 0xde714060}}}
\put(86,10){\line(1,0){10}}
\put(96,5){\oval(10,10)[tr]}
\put(101,5){\line(0,-1){20}}
\put(106,-15){\oval(10,10)[bl]}
\put(106,-20){\vector(1,0){28}}

\put(5,-2){\oval(10,10)[bl]}
\put(5,-7){\line(1,0){33}}
\put(48,-7){\line(1,0){33}}
\put(38,-12){\oval(10,10)[tr]}
\put(48,-12){\oval(10,10)[tl]}
\put(81,-2){\oval(10,10)[br]}
\put(0,-27){\makebox(86,20){\footnotesize{known to be {\tt foo\_t *}}}}
}

\put(160,10){
\put(65,130){\line(1,0){10}}
\put(75,135){\oval(10,10)[br]}
\put(80,135){\line(0,1){60}}
\put(85,200){\vector(1,0){30}}
\put(85,195){\oval(10,10)[tl]}
}

\put(160,-10){
\put(65,130){\line(1,0){30}}
\put(95,125){\oval(10,10)[tr]}
\put(100,125){\line(0,-1){15}}
\put(105,105){\vector(1,0){120}}
\put(105,110){\oval(10,10)[bl]}
}

\put(160,-30){
\put(65,130){\line(1,0){10}}
\put(75,125){\oval(10,10)[tr]}
\put(80,125){\line(0,-1){30}}
\put(85,90){\vector(1,0){30}}
\put(85,95){\oval(10,10)[bl]}
}

\put(135, 70){
\put(20, 87){\makebox(70,20)[b]{{\tt 0xde714060}}}
\put(0, 60){
\put(0, 0){\makebox(17,20)[r]{{\tt +0x0}}}
\put(20, 0){\makebox(70,20){{\tt 0xde704078}}}
}
\put(0, 40){
\put(0, 0){\makebox(17,20)[r]{{\tt +0x4}}}
\put(20, 0){\makebox(70,20){{\tt 0xde701df0}}}
}
\put(0, 20){
\put(0, 0){\makebox(17,20)[r]{{\tt +0x8}}}
\put(20, 0){\makebox(70,20){{\tt 0xdefed090}}}
}
\put(0, 0){
\put(0, 0){\makebox(17,20)[r]{{\tt +0xc}}}
\put(20, 0){\makebox(70,20){{\tt 0x12e}}}
}
\multiput(20, 20)(0, 20){3}{
\multiput(0, 0)(4, 0){18}{\line(1, 0){2}}}

\put(20, 0){\framebox(70,80){}}
}

\put(255, 130){
\put(20, 87){\makebox(70,20)[b]{{\tt 0xde704078}}}
\put(0, 60){
\put(0, 0){\makebox(17,20)[r]{{\tt +0x0}}}
\put(20, 0){\makebox(70,20){{\tt 0xde711008}}}
}
\put(0, 40){
\put(0, 0){\makebox(17,20)[r]{{\tt +0x4}}}
\put(20, 0){\makebox(70,20){{\tt 0xde70a4c8}}}
}
\put(0, 20){
\put(0, 0){\makebox(17,20)[r]{{\tt +0x8}}}
\put(20, 0){\makebox(70,20){{\tt 0xde709de0}}}
}
\put(0, 0){
\put(0, 0){\makebox(17,20)[r]{{\tt +0xc}}}
\put(20, 0){\makebox(70,20){{\tt 0xa0}}}
}
\multiput(20, 20)(0, 20){3}{
\multiput(0, 0)(4, 0){18}{\line(1, 0){2}}}

\put(20, 0){\framebox(70,80){}}
\put(90,70){
\put(0, 0){\line(1,0){20}}
\multiput(20,0)(8, 0){4}{\line(1,0){3}}
\put(52,0){\vector(1,0){10}}
}
\put(90,50){
\put(0, 0){\line(1,0){20}}
\multiput(20,0)(8, 0){4}{\line(1,0){3}}
\put(52,0){\vector(1,0){10}}
}
\put(90,30){
\put(0, 0){\line(1,0){20}}
\multiput(20,0)(8, 0){4}{\line(1,0){3}}
\put(52,0){\vector(1,0){10}}
}
}
}

\put(255, 20){
\put(20, 47){\makebox(70,20)[b]{{\tt 0xdefed090}}}
\put(0, 20){
\put(0, 0){\makebox(17,20)[r]{{\tt +0x0}}}
\put(20, 0){\makebox(70,20){{\tt 0x1}}}
}
\put(0, 0){
\put(0, 0){\makebox(17,20)[r]{{\tt +0x4}}}
\put(20, 0){\makebox(70,20){{\tt 0xde5be840}}}
}
\multiput(20, 20)(0, 20){1}{
\multiput(0, 0)(4, 0){18}{\line(1, 0){2}}}

\put(20, 0){\framebox(70,40){}}

\put(90,10){
\put(0, 0){\line(1,0){5}}
\multiput(5,0)(8, 0){3}{\line(1,0){3}}
\put(29,0){\vector(1,0){10}}
}
}

\put(365, 55){
\put(20, 47){\makebox(70,20)[b]{{\tt 0xde701df0}}}
\put(0, 20){
\put(0, 0){\makebox(17,20)[r]{{\tt +0x0}}}
\put(20, 0){\makebox(70,20){{\tt 0x646e6172}}}
}
\put(0, 0){
\put(0, 0){\makebox(17,20)[r]{{\tt +0x4}}}
\put(20, 0){\makebox(70,20){{\tt 0xba000000}}}
}
\multiput(20, 20)(0, 20){1}{
\multiput(0, 0)(4, 0){18}{\line(1, 0){2}}}

\put(20, 0){\framebox(70,40){}}
}
}

\end{picture}
\end{center}
\caption{\small{Initialization.  Each dynamically allocated
memory object is a node; each pointer between objects is an edge.  Statically
allocated objects such as {\tt foo\_list} are added to the graph as nodes.
Pointers contained within them to dynamically allocated objects are added
as edges, and the nodes themselves are marked with their type.
}
}
\label{figinit}
\end{figure*}
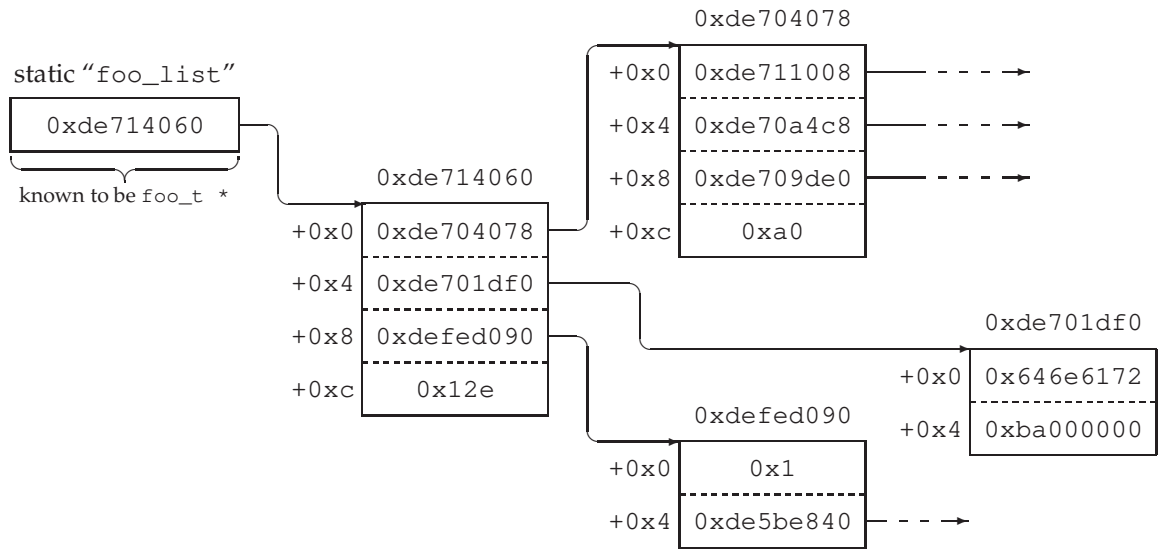

\begin{figure*}[t]
\begin{center}
\begin{picture}(465,225)

\put(-19,0){
\put(0, 0){
\put(22, 170){
\put(0, 25){\makebox(86,20)[b]{static ``{\tt foo\_list}''}}
\put(0, 0){\framebox(86,20){{\tt 0xde714060}}}
\put(86,10){\line(1,0){10}}
\put(96,5){\oval(10,10)[tr]}
\put(101,5){\line(0,-1){20}}
\put(106,-15){\oval(10,10)[bl]}
\put(106,-20){\vector(1,0){28}}

\put(5,-2){\oval(10,10)[bl]}
\put(5,-7){\line(1,0){33}}
\put(48,-7){\line(1,0){33}}
\put(38,-12){\oval(10,10)[tr]}
\put(48,-12){\oval(10,10)[tl]}
\put(81,-2){\oval(10,10)[br]}
\put(0,-27){\makebox(86,20){\footnotesize{known to be {\tt foo\_t *}}}}
}

\put(160,10){
\put(65,130){\line(1,0){10}}
\put(75,135){\oval(10,10)[br]}
\put(80,135){\line(0,1){60}}
\put(85,200){\vector(1,0){30}}
\put(85,195){\oval(10,10)[tl]}
}

\put(160,-10){
\put(65,130){\line(1,0){30}}
\put(95,125){\oval(10,10)[tr]}
\put(100,125){\line(0,-1){15}}
\put(105,105){\vector(1,0){120}}
\put(105,110){\oval(10,10)[bl]}
}

\put(160,-30){
\put(65,130){\line(1,0){10}}
\put(75,125){\oval(10,10)[tr]}
\put(80,125){\line(0,-1){30}}
\put(85,90){\vector(1,0){30}}
\put(85,95){\oval(10,10)[bl]}
}

\put(135, 70){
\put(20, 87){\makebox(70,20)[b]{{\tt 0xde714060}}}
\put(0, 60){
\put(0, 0){\makebox(17,20)[r]{{\tt +0x0}}}
\put(20, 0){\makebox(70,20){{\tt 0xde704078}}}
}
\put(0, 40){
\put(0, 0){\makebox(17,20)[r]{{\tt +0x4}}}
\put(20, 0){\makebox(70,20){{\tt 0xde701df0}}}
}
\put(0, 20){
\put(0, 0){\makebox(17,20)[r]{{\tt +0x8}}}
\put(20, 0){\makebox(70,20){{\tt 0xdefed090}}}
}
\put(0, 0){
\put(0, 0){\makebox(17,20)[r]{{\tt +0xc}}}
\put(20, 0){\makebox(70,20){{\tt 0x12e}}}
}
\multiput(20, 20)(0, 20){3}{
\multiput(0, 0)(4, 0){18}{\line(1, 0){2}}}

\put(25,-2){\oval(10,10)[bl]}
\put(25,-7){\line(1,0){25}}
\put(60,-7){\line(1,0){25}}
\put(50,-12){\oval(10,10)[tr]}
\put(60,-12){\oval(10,10)[tl]}
\put(85,-2){\oval(10,10)[br]}
\put(20, 0){\framebox(70,80){}}
\put(20,-27){\makebox(70,20){\footnotesize{inferred to be {\tt foo\_t}}}}
}

\put(255, 130){
\put(20, 87){\makebox(70,20)[b]{{\tt 0xde704078}}}
\put(0, 60){
\put(0, 0){\makebox(17,20)[r]{{\tt +0x0}}}
\put(20, 0){\makebox(70,20){{\tt 0xde711008}}}
}
\put(0, 40){
\put(0, 0){\makebox(17,20)[r]{{\tt +0x4}}}
\put(20, 0){\makebox(70,20){{\tt 0xde70a4c8}}}
}
\put(0, 20){
\put(0, 0){\makebox(17,20)[r]{{\tt +0x8}}}
\put(20, 0){\makebox(70,20){{\tt 0xde709de0}}}
}
\put(0, 0){
\put(0, 0){\makebox(17,20)[r]{{\tt +0xc}}}
\put(20, 0){\makebox(70,20){{\tt 0xa0}}}
}
\multiput(20, 20)(0, 20){3}{
\multiput(0, 0)(4, 0){18}{\line(1, 0){2}}}

\put(20, 0){\framebox(70,80){}}
\put(25,-2){\oval(10,10)[bl]}
\put(25,-7){\line(1,0){25}}
\put(60,-7){\line(1,0){25}}
\put(50,-12){\oval(10,10)[tr]}
\put(60,-12){\oval(10,10)[tl]}
\put(85,-2){\oval(10,10)[br]}
\put(20,-27){\makebox(70,20){\footnotesize{inferred to be {\tt foo\_t}}}}
\put(90,70){
\put(0, 0){\line(1,0){20}}
\multiput(20,0)(8, 0){4}{\line(1,0){3}}
\put(52,0){\vector(1,0){10}}
}
\put(90,50){
\put(0, 0){\line(1,0){20}}
\multiput(20,0)(8, 0){4}{\line(1,0){3}}
\put(52,0){\vector(1,0){10}}
}
\put(90,30){
\put(0, 0){\line(1,0){20}}
\multiput(20,0)(8, 0){4}{\line(1,0){3}}
\put(52,0){\vector(1,0){10}}
}
}
}

\put(255, 20){
\put(20, 47){\makebox(70,20)[b]{{\tt 0xdefed090}}}
\put(0, 20){
\put(0, 0){\makebox(17,20)[r]{{\tt +0x0}}}
\put(20, 0){\makebox(70,20){{\tt 0x1}}}
}
\put(0, 0){
\put(0, 0){\makebox(17,20)[r]{{\tt +0x4}}}
\put(20, 0){\makebox(70,20){{\tt 0xde5be840}}}
}
\multiput(20, 20)(0, 20){1}{
\multiput(0, 0)(4, 0){18}{\line(1, 0){2}}}

\put(20, 0){\framebox(70,40){}}

\put(25,-2){\oval(10,10)[bl]}
\put(25,-7){\line(1,0){25}}
\put(60,-7){\line(1,0){25}}
\put(50,-12){\oval(10,10)[tr]}
\put(60,-12){\oval(10,10)[tl]}
\put(85,-2){\oval(10,10)[br]}
\put(20,-27){\makebox(70,20){\footnotesize{inferred to be {\tt bar\_t}}}}
\put(90,10){
\put(0, 0){\line(1,0){5}}
\multiput(5,0)(8, 0){3}{\line(1,0){3}}
\put(29,0){\vector(1,0){10}}
}
}

\put(365, 55){
\put(20, 47){\makebox(70,20)[b]{{\tt 0xde701df0}}}
\put(0, 20){
\put(0, 0){\makebox(17,20)[r]{{\tt +0x0}}}
\put(20, 0){\makebox(70,20){{\tt 0x646e6172}}}
}
\put(0, 0){
\put(0, 0){\makebox(17,20)[r]{{\tt +0x4}}}
\put(20, 0){\makebox(70,20){{\tt 0xba000000}}}
}
\multiput(20, 20)(0, 20){1}{
\multiput(0, 0)(4, 0){18}{\line(1, 0){2}}}

\put(20, 0){\framebox(70,40){}}

\put(25,-2){\oval(10,10)[bl]}
\put(25,-7){\line(1,0){25}}
\put(60,-7){\line(1,0){25}}
\put(50,-12){\oval(10,10)[tr]}
\put(60,-12){\oval(10,10)[tl]}
\put(85,-2){\oval(10,10)[br]}
\put(20,-27){\makebox(70,20){\footnotesize{inferred to be {\tt char}}}}
}

\put(22, 0){
\put(0,107){\makebox(86,20)[b]{{\tt foo\_t} definition}}
\put(0, 80){
\put(0, 0){\makebox(34,20){{\em offset}}}
\put(38, 0){\makebox(70,20)[l]{{\em type}}}
}
\put(0, 60){
\put(0, 0){\makebox(34,20){{\tt 0x0}}}
\put(38, 0){\makebox(70,20)[l]{{\tt foo\_t *}}}
}
\put(0, 40){
\put(0, 0){\makebox(34,20){{\tt 0x4}}}
\put(38, 0){\makebox(70,20)[l]{{\tt char *}}}
}
\put(0, 20){
\put(0, 0){\makebox(34,20){{\tt 0x8}}}
\put(38, 0){\makebox(70,20)[l]{{\tt bar\_t *}}}
}
\put(0, 0){
\put(0, 0){\makebox(34,20){{\tt 0xc}}}
\put(38, 0){\makebox(70,20)[l]{{\tt int}}}
}
\put(5,0){\line(1,0){76}}
\put(86,5){\line(0,1){90}}
\put(81,100){\line(-1,0){76}}
\put(0,95){\line(0,-1){90}}
\put(5,5){\oval(10,10)[bl]}
\put(81,5){\oval(10,10)[br]}
\put(81,95){\oval(10,10)[tr]}
\put(5,95){\oval(10,10)[tl]}

\put(0,80){\line(1,0){86}}
\put(34,0){\line(0,1){100}}
\multiput(0, 20)(0, 20){3}{
\multiput(0, 0)(4, 0){22}{\line(1, 0){2}}
}
}
}

\end{picture}
\end{center}
\caption{\small{Processing.  Based on the known type of {\tt foo\_list}, we
infer that the
node it points to (the dynamic memory object at
{\tt 0xde714060}) is a {\tt foo\_t}.  Advancing to {\tt 0xde714060} and
using the type definition for {\tt foo\_t}, we determine the type of each
outgoing edge; dereferencing each type yields an inference for 
each pointed-to object.  For example, the edge at offset zero is a
pointer to a {\tt foo\_t}; we infer that the node pointed to by the
memory at offset zero ({\tt 0xde704078}) is a {\tt foo\_t}.  Likewise, 
we infer that the
object pointed to by the memory at offset four is of type {\tt char} and
that pointed to by the memory at offset eight is of type {\tt bar\_t}.
}
}
\label{figprocess}
\end{figure*}
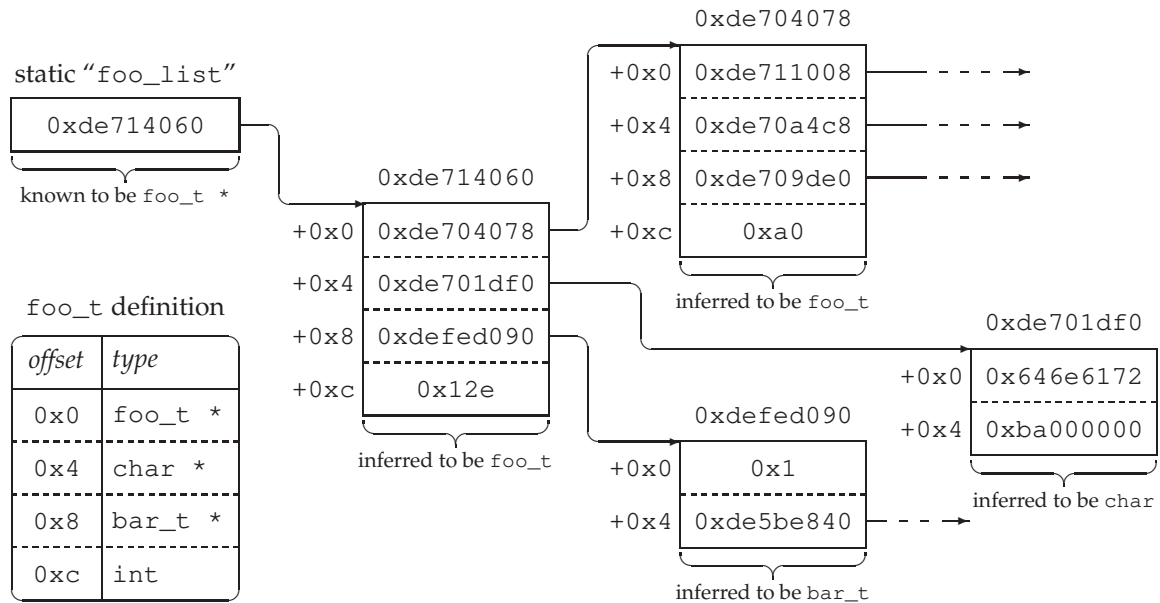

\section{Object type identification}
\label{sec-algorithm}

We seek to aid postmortem analysis by providing type identification for
dynamically-allocated objects.
In many systems, a memory dump due to fatal failure includes
both compiler-supplied type information and a mapping of static
memory objects to their type\cite{linton90evolution}; we wish to use
this information to develop type inferences for dynamically-allocated
objects.

\subsection{Initialization}
\label{sec-init}

We consume the memory dump,
building a graph in which each node represents an allocated dynamic
memory object
and each edge represents a pointer from one memory object to another. 
(We determine the pointers contained within a memory object by scanning
its aligned locations for values that correspond to other dynamically
allocated objects.)
For each node, we store the base address and size of the memory object,
as well as type information (initially set to be {\em unknown}) and a list of
outgoing and incoming edges.
For each edge, we store the offset of the pointer in the pointed-from
memory object (the {\em source offset}), and the offset pointed to in the
pointed-to memory object (the {\em destination offset}).
We then add to the graph a node for each static memory object, adding outgoing
edges as appropriate for the pointers contained in the object.  For
these static objects, we can use the type 
information generated by the compiler to set the node's type.
Figure \ref{figinit} shows an example of such a graph construction.

\subsection{Processing}
\label{sec-process}

We process the graph to propagate type information from nodes
of known type (initially, the nodes representing the static memory objects)
to nodes of unknown type.
We begin
this processing by marking and enqueuing all nodes of known type.
While the queue is non-empty, we dequeue a node and process it.
For each of the node's outgoing edges, we use the node's inferred type
and the edge's source offset within the memory object to determine the
pointer type of the edge.  If the destination offset of the edge is zero
(that is, if the represented pointer refers to the base of the pointed-to
object), we set the type of the edge's destination
node to be the dereferenced edge pointer type.  If the (now-identified)
destination node is unmarked, we mark and enqueue the destination node.
Figure \ref{figprocess} shows how this processing would proceed on the
graph from Figure \ref{figinit}.

As virtually all non-leaked dynamic memory objects are ultimately
rooted in a static memory object, this process identifies type information
for practically all dynamic memory objects.\footnote{There may 
exist some (very small) number of dynamic memory objects that are
rooted only in a thread stack or machine register;
these objects cannot be identified by this process.}
Once processing has completed,
the debugger may be queried for the type of any memory object by
providing the object's address.

\section{Complexities due to C}
\label{sec-problems}

The algorithm for postmortem type identification is straight-forward, but
several complexities arise when applying it to the C programming language
in particular.

C allows (and even encourages) type casting:
nothing stops the
programmer from storing a pointer to an object of one type as a pointer to a
different type.\footnote{C's union construct can be
thought of as a slightly more sanctioned variant of type casting:  instead
of casting to arbitrary type, the compiler enforces that a given datum
may only be interpreted as one of a specified list of types.}
If our algorithm encountered such a pointer, it would
incorrectly assign the destination node to be the cast-to type.  Worse,
the node would be {\em propagated} as the incorrect type, potentially
leading to more misidentifications and mispropagations.  
In theory, the presence of type casting in C merely prevents us from
guaranteeing the correctness of type inference; in practice, its ubiquity
reduces our algorithm to a series of heuristics.  

One highly effective heuristic may be to {\em only} propagate type
information when the size of the pointed-to type equals the size of the
pointed-to object.  This would limit incorrect behavior to cases where the
programmer is storing a pointer to an object of one type as a pointer to a
different type of {\em identical} size.  However, this would also limit
propagation unreasonably: even in well-written C, it is {\em often} the case
that pointed-to objects are not the size of the pointing type.  These
objects can occur for many reasons, but several phenomena --- discussed
below --- seem to be
responsible for most examples in practice.

\subsection{Array declaration syntax}

Most ubiquitous of these phenomena is C's array declaration syntax.
Unlike Pascal and other languages, C makes no distinction between the
declaration of a pointer to a single object and a pointer to an array of
objects of like type.  That is, in the declaration

\small
\begin{verbatim}
          struct foo *bar;
\end{verbatim}
\normalsize
``{\tt bar}'' could be a pointer to a single {\tt foo} structure, or it could
point to an array of {\tt foo} structures.  

\subsection{Flexible array members}

C performs no bounds checking on array indexing, allowing declarations
of structures that are implicitly followed by arrays of the type of the
structure's last member.  For example, in the declaration

\small
\begin{verbatim}
 	    typedef struct foo {
 	            int       foo_bar;
 	            int       foo_baz;
 	            mumble_t  foo_mumble[1];
 	    } foo_t;
\end{verbatim}
\normalsize
when the programmer allocates a {\tt foo\_t}, the size of $n - 1$
{\tt mumble\_t}s is added to the size of a {\tt foo\_t} to derive the 
total size of the allocation; this
allows for the $n$ trailing {\tt mumble\_t}s to be referenced from the
allocated
{\tt foo\_t} using C's convenient array syntax --- and without requiring an
additional
memory dereference.  This technique may seem arcane, but it is 
so widespread that ISO C99 has a name for the last member in such a structure:
the ``flexible array member'' (FAM)\cite{ISO:1999:IIP}.

\subsection{Structure embedding}
\label{sec-embed}

In object-oriented systems implemented in C, it is common for structures to
contain embedded smaller
structures and to pass pointers to these smaller structures to routines
that track
the structures only by the smaller, embedded type.  This effects a
crude polymorphism:  the larger structure inherits the data
of the smaller structure and methods can be called on that data by specifying
a pointer to the embedded structure as a first argument.
This technique has been described at some length by Siff et
al.\cite{siff99coping}, and examples of it abound.
In the Solaris kernel the most pervasive example
is the virtual filesystem: file systems
typically define their own file system node type that embeds the 
general system's virtual node type, ``{\tt vnode\_t}''\cite{Kleiman:1986:VAM}.
The virtual
file system contains data structures of {\tt vnode\_t}s, but each is
actually the embedded {\tt vnode\_t} in a larger, file system-specific
type.

It is less common (but by no means unheard of) for smaller structures to
be used as place holders in data structures consisting of a larger
structure.  That is, instead of an instance of a larger structure being
pointed to by the smaller structure pointer (as was the case with structures
of {\tt vnode\_t}s, above), an instance of the smaller
structure is pointed to by the larger structure pointer.  This technique is
somewhat dubious, but is most often used to implement hash tables in which a
circular linked list of table elements is desired.  In this implementation, the
smaller structure might be declared this way:

\small
\begin{verbatim}
      typedef struct fooent {
              struct foo  *foo_prev;
              struct foo  *foo_next;
              int          foo_flags;
      } fooent_t;
\end{verbatim}
\normalsize

The first members of the larger structure will be identical to that of the
smaller structure, but the structure will contain additional data, e.g.:

\small
\begin{verbatim}
      typedef struct foo {
              struct foo  *foo_prev;
              struct foo  *foo_next;
              int         foo_flags;
              struct bar  foo_bar;
      } foo_t;
\end{verbatim}
\normalsize

The hash table itself will be a table of {\tt fooent\_t} instead of
{\tt foo\_t} ---
thereby saving space in the table itself while still allowing circular lists
of {\tt foo\_t} structures.
This construct is critically important to identify:
it often occurs in large, contiguous arrays of
the smaller structure; mispropagating these as arrays of the larger
structure would result in wide-spread misidentification.

\section{Heuristics}
\label{sec-heuristics}

We have developed a series of heuristics to implement the algorithm
described in Section \ref{sec-algorithm} while mitigating the inherent
difficulties presented in Section \ref{sec-problems}.  Wherever possible,
these heuristics attempt to avoid
mispropagation: no identification is preferred to misidentification.

\subsection{Conservative propagation}
\label{sec-conservative}

We initialize the graph as described in Section~\ref{sec-init}, but
instead of each node having a single type identifier, each stores
a {\em list} of possible types.  We propagate types out from the known
nodes as described in Section \ref{sec-process}, but proceeding
as {\em conservatively} as possible.  Specifically:

\begin{itemize}
\item
We do not propagate an inferred type if the size of the type is less
than twice the size of the object.  These nodes may be examples of the
phenomena
described in Section~\ref{sec-problems}; they are specifically addressed by
other heuristics.

\item We do not propagate if the inferred type is a union.

\item If we discover a new type inference for a node, we add it to
the node's type list --- but we only propagate through the node 
if it is unmarked.  (The node is marked as we propagate through it.)
This prevents any node from being propagated with multiple, different
inferences.

\item If the destination offset is something other than zero (that is, if the
edge does not point to the base of an object), we propagate based on
the
destination type but we do {\em not} add the type to the destination
node's type list. This allows us to conservatively propagate 
through embedded types.
\end{itemize}

\subsection{Embedded type detection}

Embedded types are largely dealt with by virtue of conservative propagation:
because type inferences are only added when an edge points to the base
of a structure, we can only potentially misinterpret a node to be its embedded
type
if the embedded type is the first member of the encapsulating structure.
If this is the case, conservative propagation will hopefully yield
{\em multiple inferences} for the node: because we refuse to perform
further processing on any node that has multiple inferences, this
will prevent a node with an embedded type as its first member
from being misinterpreted as an array of the embedded type.

\subsection{FAM detection}

To detect FAMs, we visit each 
node for which we have exactly one inferred type, and for which the size
of the object is greater than or equal to twice the size of the type.  
(These are nodes for which we made a type inference but through which
we refused to propagate during conservative propagation.) 
To differentiate an array of the inferred type from a type with a flexible
array member, we resort to an inelegant but effective technique:  
we check the last member
of the inferred type; if the structure ends with an array of size one, it is
deemed to have a flexible array member and processing advances to 
array propagation.  This assumes that the {\em only} reason that one would
have the last member of a structure be an array of size one is to use it as
a flexible array member.  While one can clearly develop counterexamples, this
technique works well in practice --- and the cautious array propagation
described in Section~\ref{sec-arrayprop} prevents mispropagation should
a counterexample be encountered in the wild.\footnote{ISO C99 defines
an alternate syntax to denote a FAM:  instead of declaring
{\tt foo\_mumble[1]}, one may declare {\tt foo\_mumble[]}.  Assuming
that this information percolates into the
compiler-supplied type information, this will allow for reliable 
FAM detection.}

\begin{figure}[t!]
\begin{center}
\begin{picture}(227,168)

\put(40,84){\makebox(0, 0)[r]{\small{224 bytes}}}

\put(54,5){\oval(10, 10)[bl]}
\put(54,163){\oval(10, 10)[tl]}
\put(49,5){\line(0,1){74}}
\put(44,79){\oval(10, 10)[tr]}
\put(44,89){\oval(10, 10)[br]}
\put(49,89){\line(0,1){74}}

\put(56,0){
\put(0,0){\framebox(72,168){}}
\multiput(0,54)(5, 0){15}{\line(1, 0){2}}
\put(74,5){\oval(10, 10)[br]}
\put(74,49){\oval(10, 10)[tr]}
\put(84,32){\oval(10, 10)[bl]}
\put(84,22){\oval(10, 10)[tl]}
\put(79,5){\line(0,1){17}}
\put(79,32){\line(0,1){17}}
\put(88,27){\makebox(0, 0)[l]{\small{72 bytes}}}

\multiput(0,111)(5, 0){15}{\line(1, 0){2}}
\multiput(72,54)(8, 0){6}{\line(1, 0){1}}
\multiput(72,168)(8, 0){6}{\line(1, 0){1}}

\put(0,54){
\put(74,5){\oval(10, 10)[br]}
\put(74,52){\oval(10, 10)[tr]}
\put(84,33.5){\oval(10, 10)[bl]}
\put(84,23.5){\oval(10, 10)[tl]}
\put(79,5){\line(0,1){19}}
\put(79,33.5){\line(0,1){19}}
\put(87,28.5){\makebox(0, 0)[l]{\small{76 bytes}}}
}

\put(0, 0){\makebox(72,54){\small{\em Remainder}}}
\put(0, 54){\makebox(72,57){\small{\em Element 1}}}
\put(0, 111){\makebox(72,57){\small{\em Element 0}}}

\put(0,111){
\put(74,5){\oval(10, 10)[br]}
\put(74,52){\oval(10, 10)[tr]}
\put(84,33.5){\oval(10, 10)[bl]}
\put(84,23.5){\oval(10, 10)[tl]}
\put(79,5){\line(0,1){19}}
\put(79,33.5){\line(0,1){19}}
\put(87,28.5){\makebox(0, 0)[l]{\small{76 bytes}}}
}

\put(119,163){\oval(10, 10)[tr]}
\put(119,59){\oval(10, 10)[br]}
\put(124,59){\line(0,1){47}}
\put(129,106){\oval(10, 10)[tl]}
\put(129,116){\oval(10, 10)[bl]}
\put(124,116){\line(0,1){47}}
\put(133,111){\makebox(0, 0)[l]{\small{152 bytes}}}
}
\end{picture}
\caption{\small{
Evaluating a memory object as an array of an inferred type.
In this example, the size of the inferred type is 76 bytes and the size of
the memory object is 224 bytes.  Were this an array of the inferred type,
it would contain only
two elements --- there is not enough space in 224 bytes to fit a third.  A
two element array would be 152 bytes in size; if there exists a
general-purpose object cache with objects smaller
than 224 bytes but greater than or equal to 152 bytes, we will conclude that
this is {\em not} an array of the inferred type.}}
\label{figarray}
\end{center}
\end{figure}
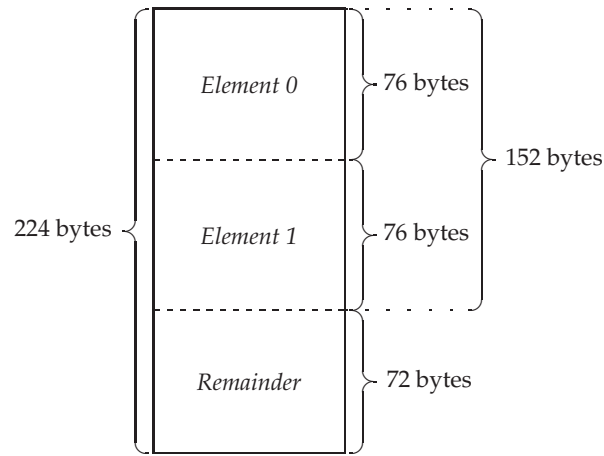

\subsection{Array determination}
\label{sec-array}

If a node is deemed to not have a FAM, we must determine if it is an array
of the inferred type.  To do this, we make an important check that requires
some explanation of the object-caching memory allocator used by the Solaris
kernel\cite{bonwick94slab}.  In this allocator, objects may be allocated
either by specifying an object-specific cache, or they may be
allocated by specifying the amount of memory desired.  Because establishing an
object-specific cache presents an additional complexity for the programmer,
most objects are allocated by just specifying the object size.  As
all objects are allocated out of {\em some} cache, 
a number of general-purpose object caches are created by the
allocator itself, with each cache corresponding to a fixed, common size.
The allocator supports size-based allocations by allocating out
of the general-purpose cache with the smallest object size that will
satisfy the request.

Because all objects from a given cache are of fixed size, all
dynamic arrays are allocated out of a general-purpose cache.
We can use this
implementation detail to perform an additional check on any
potential array: we calculate the size of the object modulo the size of
the inferred type and subtract it from the size of the object.
If this value is less than or
equal to the object size of the next-smaller general-purpose cache, we
know that
it is {\em not} an array of the inferred type --- if it were an array of the
inferred type, it would have been instead allocated out of the next-smaller
cache. This determination is shown in Figure \ref{figarray}.  

While we have couched our array determination technique in terms of the 
Solaris kernel memory allocator, it is actually applicable to
{\em any} memory allocator.  The efficacy of the technique will vary
depending on the degree to which the size of an allocated buffer matches
the size of the allocation request:  the technique will be most effective
on allocators that exactly match allocated buffer size to requested buffer
size, or otherwise track requested buffer size on a per-buffer basis.

\subsection{Array propagation}
\label{sec-arrayprop}

In the case that an array or a FAM is detected, the type of each element
of the array must be propagated.
Because the array and FAM determination heuristics are imperfect,
array propagation runs the risk of propagating incorrect types.
To mitigate this risk, we perform an additional check before propagating
an array:
we iterate through each element of the
hypothesized array, checking that each pointer member points to either NULL
or valid memory.  If pointer members do not satisfy these criteria, it is
assumed that we have not accurately determined that the given object is
an array of the inferred type, and we abort processing of the node.  Note
that uninitialized pointers can potentially prevent an otherwise valid
array from being interpreted as such.  Array misinterpretation
can induce substantial cascading type misinterpretation; it is preferred to
be conservative and accurate in such cases --- even if it means a lower type
recognition rate.  An array is propagated by propagating the type of each
array element using conservative propagation.

\section{Implementation}
\label{sec-implementation}

We have implemented postmortem type identification as a debugger command
in Solaris's modular debugger, MDB\cite{mdb}.  MDB provides an API that
allows 
for the rapid development of pluggable debugging components, and includes
a specific API for the processing of the type information
present in all Solaris kernel memory dumps.  MDB's architecture allows
new components to easily build on extant ones, trivializing otherwise
complex tasks such as iterating over all allocated memory objects.

Type identification is performed by using the ``{\tt ::typegraph}''
debugger command.  To avoid propagating incorrect type information if
at all possible, {\tt ::typegraph} applies our heuristics in a series
of passes, with more aggressive heuristics applied to only those
nodes for which more conservative heuristics have failed to make a type
identification.

\begin{figure}[b!]
\begin{center}
\footnotesize
\begin{verbatim}
             typegraph:                           pass => initial
             typegraph:                  maximum nodes => 1401799
             typegraph:                   actual nodes => 824313
             typegraph:                 anchored nodes => 4992
             typegraph:        time elapsed, this pass => 1526 seconds
             typegraph:            time elapsed, total => 1526 seconds
             typegraph:
             typegraph:                           pass => 1
             typegraph:                          nodes => 824313
             typegraph:                       unmarked => 576829        (69.9%)
             typegraph:                          known => 255955        (31.0%)
             typegraph:                    conjectured => 242699        (29.4%)
             typegraph:          conjectured fragments => 3737          ( 0.4%)
             typegraph:           known or conjectured => 502391        (60.9%)
             typegraph:                      conflicts => 35878
             typegraph:                     candidates => 172816
             typegraph:        time elapsed, this pass => 44 seconds
             typegraph:            time elapsed, total => 1570 seconds
             typegraph:
             typegraph:                           pass => 2
             typegraph:                          nodes => 824313
             typegraph:                       unmarked => 275390        (33.4%)
             typegraph:                          known => 255955        (31.0%)
             typegraph:                    conjectured => 535919        (65.0%)
             typegraph:          conjectured fragments => 4049          ( 0.4%)
             typegraph:           known or conjectured => 795923        (96.5%)
             typegraph:                      conflicts => 39132
             typegraph:                     candidates => 15597
             typegraph:        time elapsed, this pass => 81 seconds
             typegraph:            time elapsed, total => 1652 seconds
             typegraph:
\end{verbatim}
\normalsize
\end{center}
\normalsize
\caption{\small{{\tt ::typegraph} output from the
first two passes (conservative
propagation and array determination) for a dump with 824,313 dynamic
memory objects.  First,
note that a relatively large number of objects
(31.0\%) are known after the initial pass.  This is due to the known 
kernel object caches.  Second, note that while conservative propagation 
identifies a large number of objects (29.4\%), there are still many
objects (39.1\%) unidentified after the first pass.  The array determination
pass is critical for identifying these objects; after this pass, types are
known or conjectured for 96.5\% of objects.  
These results are typical.
Finally, note that much more time is spent in the initial pass than in
either of the subsequent passes.  This is because the initial pass includes
the time to read the crash dump from disk into memory --- and the 
crash dump is over two gigabytes in this example.
}}
\end{figure}

\subsection{Initial pass}

The initial pass uses the internal data structures of the Solaris
kernel memory
allocator to iterate over all allocated dynamic memory objects and build
the graph as described in Sections \ref{sec-init} and \ref{sec-conservative}.
Our implementation adds an important additional step to the initialization:
because the Solaris
kernel uses an object-based allocator, we can iterate over nodes from
kernel memory caches of known type and
set their type accordingly.  (For example, the objects allocated from
the ``{\tt process\_cache}'' are known to be of type ``{\tt proc\_t}.'')
This technique requires very little encoded
knowledge of the system, but allows for substantial preprocessing
identification:  our current implementation contains a table consisting of
only nine cache/type pairs, but it leads to {\em a priori}
identification of up to a third of all dynamic memory objects.

\subsection{Processing passes}

Our heuristics vary in the assumptions they make.  To avoid mispropagation,
we process the graph as a series of passes, applying more aggressive
heuristics only to those nodes for which conservative heuristics have
failed to make an identification.

\subsubsection{Conservative propagation}  
Conservative propagation is the first processing pass, proceeding exactly
as described in Section \ref{sec-conservative}.

\subsubsection{Array determination}  
The array determination pass processes all nodes through which we did
not propagate in the first pass, proceeding exactly as described in
Section \ref{sec-array}.
If a node is determined to be an array or is determined to have a FAM,
the array is conservatively propagated as described in
Section \ref{sec-arrayprop}.
Because this propagation can lead to previously unidentified nodes being
identified as potential arrays, this pass is 
repeated until no improvements are made.

\subsubsection{Type coalescence} 
The type coalescence pass processes all nodes that have multiple type
inferences.  If a node has one inference that is a structure and
others that are not a structure, the non-structure inferences are 
eliminated.  For example if an object is inferred to be either of type
``{\tt char}'' (pointed to by a ``{\tt char *}'') or type 
``{\tt struct frotz},'' the possibilities will be coalesced into just
``{\tt struct frotz}.''

\subsubsection{Non-array type inference}  
The non-array type inference pass is the least conservative.
It processes those objects that meet the following conditions:

\begin{itemize}
\item The object has a single type inference.
\item The size of the type inference is less than half the object size.
\item The object was not identified as an array.
\end{itemize}

These objects --- which are not propagated by earlier passes --- are
propagated in this pass as being a non-array of the inferred type.
This pass is necessary to propagate objects that have embedded types as
their first member, but for which only the embedded type was inferred.

\begin{figure}[t]
\begin{center}
\footnotesize
\begin{verbatim}
         > 33a31007088::whattype
         33a31007088 is 33a31007088+0, struct seg

         > 3039b042370::whattype               
         3039b042370 is 3039b042370+0, possibly struct dcentry 

         > 30062034c3c::whattype
         30062034c3c is 30062034c38+4, possibly char (struct rnode.r_path) 

         > 329642d7878::whattype
         329642d7878 is 329642d7878+0, possibly one of the following:
           struct sonode (from 15925e0+8, type struct socklist)
           struct vnode (from 30985f1d038+10, type struct stdata)

         > 3020a383880::whattype
         3020a383880 is 3020a383880+0, possibly one of the following:
           struct filock (from 33cf55b2a80+80, type struct tmpnode)
           struct lock_descriptor (from 4823ff82a00+8, type struct lock_descriptor)
\end{verbatim}
\normalsize
\end{center}
\caption{\small{Example {\tt ::whattype} output for several different objects.
In the first example, the object in question is from a cache of
known type.  In the second example, the type of the object has
been inferred (and {\tt ::whattype} softens its language accordingly by
only claiming that this is ``possibly'' the type).  In the third
example, the type is inferred to be a base type, so {\tt ::whattype}
provides the referring type name and member.  In the final two examples,
the objects have a type conflict.  The first of these 
is an example of
the idiom discussed in Section \ref{sec-embed}:
the embedded type (``{\tt struct vnode}'') is the first member of
its encapsulating type (``{\tt struct sonode}'').  The second is simply
a result of sloppy programming:  an explicit cast has been used to store
a ``{\tt struct lock\_descriptor}'' object in a pointer to
``{\tt struct filock}'' --- despite the fact that the two structures have
nothing to do with one another!  Fortunately, such conflicts are rare;
typically fewer than 0.1\% of objects are identified as having type conflicts.
}}
\label{figwhattype}
\end{figure}

\subsection{Type identification}

Once processing is complete, the type of an arbitrary
object may be queried by specifying its address to the
``{\tt ::whattype}'' debugger command.
If an inference has been drawn for the specified object, its type
is displayed. 
If the type has been identified to be a base type (e.g., {\tt char},
{\tt int}, {\tt void}, etc.), the type and member of the referring
object is displayed.  If multiple inferences have been drawn for
the object, all of the conflicting types are displayed, along with
the nodes that led to each inference.  Figure \ref{figwhattype}
shows some example output from {\tt ::whattype}.

\subsection{Manual intervention}

If recognition rate is low (or, from a more practical perspective, if the
object of interest is not automatically identified), it can be useful
to know which node is the greatest impediment to further recognition.
If the user can --- by hook or by crook --- determine the true type of this
node,
more type identification should be
possible.  To facilitate this, we therefore define the {\em reach} of a node
to be the number of unknown nodes that could potentially be identified
were the
node's type known.  We determine the reach by performing a
depth-first pass through the graph.  The node of greatest reach (along with
the number of reachable unknown nodes)
is reported upon completion of a post-processing pass.
We added an additional debugger command, {\tt ::istype}, to allow the
type to be set manually.  When a type is set manually, the graph is
immediately reprocessed.  

Manual intervention in the presence of imperfect heuristics allows for
a paradigm of
{\em feedback-based} postmortem debugging, where automatic inferences by
the debugger about the system lead to further inferences by the user about
the system --- which in turn lead to more automatic inferences and so on.
While this is intriguing in principle, we have found that
the high rate of type recognition has not necessitated its use in
practice.

\section{Other applications}

While debugging buffer overrun corruption was our original motivating
application for postmortem type identification, we have discovered it
to have a wide range of applications to postmortem debugging.

\label{sec-otherapps}

\subsection{Use-after-free corruption}

Postmortem type identification may be of help in debugging other
variants of memory corruption.  In particular, it may help root-cause
{\em use-after-free} corruption, in which a memory object is used 
after it is deallocated.  After buffer overrun
corruption, use-after-free corruption has been found to be the most
common type of memory corruption\cite{sullivan91software}.
This pathology can be difficult to diagnose:  it manifests itself
as ``random'' corruption.  If the object as reallocated is of different type
than the object as erroneously used after being freed --- and if the
freed object is still present in its original data structure --- postmortem
type identification will explicitly identify the object as a type conflict.
By providing the two types in conflict, postmortem type identification
considerably focuses the search for errant code.

\subsection{Postmortem lock detection}

When debugging parallel software systems postmortem, it is often useful to
know which mutually exclusive regions a given thread of control has access to.
Such regions are entered by acquiring a mutual exclusion lock (``mutex''); the
problem distills to determining which mutexes are owned in the system and by
whom.  Knowing the regions that a thread of control has exclusive access
to sheds light on the state of the system at the time of the failure, and
therefore aids analysis into the failure's root cause.

The simplest way to solve this problem is to log each entry to and exit from a
mutually exclusive region.  However, in parallel software systems with
fine-grained locking, mutexes are acquired and released far too frequently to
allow for any sort of logging without inducing an unacceptable impact on
performance.

If mutexes are implemented with a single C type (as they are in the 
Solaris kernel ---
a ``{\tt kmutex\_t}'') we can build on postmortem type identification to
find all held locks:  after type identification has completed, we iterate
over all nodes of inferred type, and look for any embedded mutex types.
Adding the offset of the embedded mutex type to the base address of the
node yields the address of the mutex.  Because parallel systems must be able
to determine the owning thread of a mutex given its address (to
correctly implement priority inheritance, guard against recursive entry,
allow for adaptive blocking behavior, etc.), we can build a system-specific
way to get from the mutex to the owning thread.  

We have implemented an additional MDB command, ``{\tt ::findlocks},''
that implements this for the Solaris kernel; its output is shown
in Figure~\ref{figfindlocks}.

\begin{figure}[t!]
\begin{center}
\footnotesize
\begin{verbatim}
              > ::findlocks
              300182d9a78 (struct segkp_data.kp_lock) is owned by 140a000
              3001c69ca80 (struct anon_map.serial_lock) is owned by 3008de45a40
              3000d2c46f0 is owned by 3008de45a40
              3000c43b658 (struct anon_map.serial_lock) is owned by 30018a7bce0
              3001aabdca8 (struct anon_map.serial_lock) is owned by 30018a7a2a0
              30011b8c390 (struct anon_map.serial_lock) is owned by 30011bb6d20
              30011b8c898 (struct anon_map.serial_lock) is owned by 30018a73d00
              30011b8c940 (struct anon_map.serial_lock) is owned by 30018a7ad20
              3008423e740 (struct anon_map.serial_lock) is owned by 30078e06d20
              30320df6118 (struct anon_map.serial_lock) is owned by 30084296800
              300841be820 (struct anon_map.serial_lock) is owned by 3001ccfd260
              30320f5dcf0 (struct anon_map.serial_lock) is owned by 3000c442fe0
              14b4a18 (pageout_mutex) is owned by 2a100527d40
              14cc7a8 (ufs_scan_lock) is owned by 3000c4437c0
\end{verbatim}
\normalsize
\end{center}
\caption{\small{Example {\tt ::findlocks} output.  The address of 
the lock is shown on the left; the address of the owning thread structure
is shown on the right.
For static locks (such as ``{\tt pageout\_mutex}'' and
``{\tt ufs\_scan\_lock}'' in
the above), the symbol name is provided.  For locks embedded in dynamic
objects, the structure type of the object and member name of the 
embedded lock are provided.  {\tt ::findlocks} output should be taken
only to be advisory; if type recognition is anything less than 100\%,
it will not find all held locks.
}}
\label{figfindlocks}
\end{figure}

\subsection{False sharing detection}

In caching SMP systems, memory is kept coherent through a variety of
different protocols.  Typically, these protocols dictate that
only a single cache may
have a given line of memory in a dirty state.
If a different cache wishes to write to the
dirty line, the new cache must first read-to-own the dirty line from the
owning cache.  The size of the line used for coherence (the coherence
granularity) has an immediate ramification for parallel software:  because
only one cache may own a line at a given time, one wishes to avoid a
situation where two or more small, disjoint data structures are both
contained within a single line {\em and} accessed in parallel on disjoint
CPUs.  This situation --- so-called
{\em false sharing}\cite{dubois93detection} --- can induce
suboptimal scalability in otherwise scalable software.

\begin{figure}[t]
\begin{center}
\footnotesize
\begin{verbatim}
         > ::findfalse
                ADDR SYMBOL                       TYPE                   SZ TOTSIZE
            7839fc30 fx_list_lock                 struct mutex            8     128
            7839f6b0 fx_cb_list_lock              struct mutex            8     128
            78360448 t_hashmutex                  struct mutex            8     512
            782e2418 kcpc_ctx_llock               struct mutex            8     512
             14c3438 tcp_bind_fanout              struct tbf_s           16    8192
             14c2438 tcp_listen_fanout            struct tf_s            16    4096
             14c1428 tcp_acceptor_fanout          struct tf_s            16    4096
             14bd2e8 ipc_tcp_listen_fanout_v6     struct icf_s           16    4096
             14bb628 tbftable                     struct tbf             56    1792
         30000c07e80 -                            struct fifolock        32     288
         30000c07d60 -                            struct fifolock        32     288
         30000c07c40 -                            struct fifolock        32     288
         30000b70000 -                            struct icf_s           16    8192
         30000b6c000 -                            struct icf_s           16    8192
         30000b62000 -                            struct tbf_s           16    8192
         300008d1000 -                            struct mutex            8    4096
         30004153530 -                            struct uf_entry        40    2688
         3000284eaa8 -                            struct uf_entry        40    2688
         3000284e028 -                            struct uf_entry        40    2688
         30004149538 -                            struct uf_entry        40    2688
         30004148038 -                            struct uf_entry        40    2688
         30006cc7540 -                            struct uf_entry        40    2688
         30000c01520 -                            struct uf_entry        40    2688
         30000c00aa0 -                            struct uf_entry        40    2688
         30000c00020 -                            struct uf_entry        40    2688
         3000256f9c0 -                            struct uf_entry        40    1312
         3000256f4a0 -                            struct uf_entry        40    1312
         30000061988 -                            struct smfree          24     192
         30000ca0000 -                            struct irb             32   65536
         30000ae6000 -                            struct mutex            8  262144
         30000920000 -                            struct plock            8  124624
         30000916000 -                            struct _kcondvar        2   31156
\end{verbatim}
\normalsize
\end{center}
\caption{\small{Example of {\tt ::findfalse} output.  All the above
structures can potentially suffer from false sharing.  One of these
(``{\tt struct uf\_entry}'') was deemed important enough to fix
immediately.  Others are either under investigation or are in data
structures that are not sufficiently parallel to merit eliminating
the false sharing.
}}
\label{figfindfalse}
\end{figure}

Historically, one has been able to find false sharing only with some
combination of keen intuition and good luck.  Building on postmortem
type information we can --- from a system
crash dump --- detect the potentially most egregious cases of false
sharing.  This pushes postmortem analysis into an
entirely new domain:  analyzing a system crash dump
for potential (but as of yet unknown) performance problems.

We can detect false sharing by iterating over all nodes, looking for nodes
that satisfy the following criteria:

\begin{itemize}
\item The node is an array.  That is, the node was either determined to
        be of a C type that is an array type, or the node was inferred to be
        an array in the array determination pass of type identification.

\item Each element of the array is a structure that is smaller than the
        coherence granularity.

\item The total size of the array is greater than the coherence granularity.

\item Each element of the array is a structure that contains within it a
        synchronization primitive (mutex, readers/writer lock, condition
        variable or semaphore).  We use the presence of a synchronization
        primitive as a crude indicator that the disjoint elements of the
        array are accessed in parallel.
\end{itemize}

Any node satisfying these criteria is identified as an object that could
potentially suffer from false sharing, and the node's address,
type, type size, and total size are provided as output.
We have implemented this as a ``{\tt ::findfalse}'' debugger command;
its output is shown in Figure~\ref{figfindfalse}.

While there are some instances of false sharing that do not meet the
above criteria (e.g., if the synchronization for each element is handled
in a separate structure, or if the elements are only manipulated with
atomic memory operations), these criteria yield many examples of false
sharing without swamping the user with false positives.
(As a proof of concept, {\tt ::findfalse} has found
several known instances of false sharing in the Solaris kernel, and further
revealed two serious --- and hitherto unknown --- instances of false
sharing.)

\section{Future work}
\label{sec-future}

\subsection{Better recognition}

In our experience, the greatest impediments to type recognition are
data structures that are stored {\em only} with pointers to
base types that are subsequently cast before every use.  This may seem
arcane, but it comes up frequently in the Solaris kernel:
device driver instances register an object that represents their state with a
framework that stores it as a ``{\tt void *},'' handing it back to the
device driver as needed.  Because these objects are never stored with
a pointer to the true type, we cannot identify their type.  We would
ideally like to extend postmortem type identification to be able to identify
these structures, perhaps by extending the interfaces that create
such state to explicitly specify type.

\subsection{User-level core files}

Postmortem type identification has proved very useful for debugging Solaris
kernel crash dumps; a logical extension is to allow for type identification
in user-level core files.  
The work required to do this is relatively small:
MDB can already process user-level core files and 
the slab allocator used in the Solaris kernel has recently
been made available to user-level processes\cite{USENIX01*15}.
The only impediment is the addition of the type information consumed 
by MDB to user-level core files;
this work is currently underway.  Furthermore, because multithreaded 
applications on Solaris are forced to use well-defined types for 
synchronization primitives, we expect to be able to provide {\tt ::findlocks}
and {\tt ::findfalse} for user-level core files as well.

\section{Conclusions}

We have described a mechanism to provide automatic postmortem
identification of arbitrary memory objects.  While the technique can in
principle be applied to any typed language, we have focused on the specific
issues presented by C.
We have described heuristics to overcome the obstacles inherent in 
postmortem type identification for C-based systems, and have described our
implementation on one such system, the Solaris kernel.  We have found
postmortem type identification to be very useful in debugging buffer
overrun memory corruption in optimized systems --- problems that were
practically undebuggable prior to this work.  Moreover, we have found
that our heuristics yield a sufficiently high recognition rate to allow
for additional novel applications, including postmortem identification
of held locks and postmortem identification of structures that may
induce false sharing.  There will certainly be more applications of
postmortem object type identification, some of
which will presumably rely on near-perfect type recognition; there
is therefore great incentive to develop further heuristics to improve
the object recognition rate as much as possible.

\subsection*{Availability}

\begin{sloppypar}
The debugger commands described in this work ---
{\tt ::typegraph},
{\tt ::whattype},
{\tt ::istype},
{\tt ::findlocks} and
{\tt ::findfalse} --- have been integrated into Solaris, and will 
be available in the October 2003 update of Solaris 9.  More information
on the availability of Solaris can be found at
{\href{http://www.sun.com/software/solaris}
{\tt http://www.sun.com/software/solaris/}}.
\end{sloppypar}

\subsection*{Acknowledgements}

Mike Shapiro developed both MDB and the infrastructure to include
compiler-produced type information in Solaris kernel crash dumps; this
work would not have been possible without his tireless efforts.
Adam Leventhal extended the MDB module API to allow processing of
type information, and engaged in many useful conversations about
postmortem type identification.
Several of my colleagues in Solaris Kernel Development
reviewed drafts of this paper; it was much improved by the detailed
comments of
Jonathan Adams, 
Matt Ahrens,
Adam Leventhal,
Peter Memishian,
Dave Powell,
Daniel Price,
Mike Shapiro,
Bart Smaalders,
and Andy Tucker.
Finally, I am indebted to Brigid Gaffikin, who agreed to apply
her knowledge of punctuation arcana to editing this work.  However ---
and as she is quick to point out --- any remaining punctuation errors
are my own.

\bibliography{tg}

\end{document}